**Title:** Truncation by death in the sufficient cause framework

**Authors:** Bronner P. Gonçalves[1], Eiji Yamamoto[2], Etsuji Suzuki[3]

**Affiliations:**

[1] Faculty of Health and Medical Sciences, University of Surrey, Guildford, United Kingdom

[2] Okayama University of Science, Okayama, Japan

[3] Department of Epidemiology, Graduate School of Medicine, Dentistry and Pharmaceutical Sciences, Okayama University, Okayama, Japan

**Correspondence:**

Bronner P. Gonçalves, Faculty of Health and Medical Sciences, University of Surrey, Guildford, United Kingdom (bronnergoncalves@gmail.com)

Etsuji Suzuki, Department of Epidemiology, Graduate School of Medicine, Dentistry and Pharmaceutical Sciences, Okayama University, 2-5-1 Shikata-cho, Kita-ku, Okayama 700-8558, Japan (etsuji-s@cc.okayama-u.ac.jp)

**Abstract**

The sufficient cause framework has been used for decades to improve our understanding of both basic and more complex causal concepts in epidemiology, such as mediation and interaction. Here, we make use of this framework to provide a description of truncation by death, in which the outcome of interest is undefined for individuals who die before the time of assessment at the end of follow-up. We explain the non-causal nature of the crude estimand that compares outcomes by treatment levels conditional on observed survival by showing that it corresponds to a comparison of distinct risk status types, which are defined based on the susceptibility to sufficient causes. Further, expressions for the crude estimand and for the survivor average causal effect, a causal estimand defined under the principal stratification approach, are provided in terms of population-level joint frequencies of the background factors of sufficient causes. Finally, we also describe conditions, based on background factors of sufficient causes, under which the survivor average causal effect is null. Our description of this problem, which studies truncation by death from a new perspective, might encourage further analyses of principal stratification-based estimands using sufficient causes.

**Manuscript**

*Background*

In epidemiologic studies where the outcome of interest is only defined for individuals who are alive at the end of the follow-up, when assessment occurs, if some participants die during the follow-up, the inferential problem of truncation by death might be present [1, 2], and can affect not only observational studies but also randomised clinical trials [3] (see ICH E9(R1) Addendum [4]). In this situation, analyses that condition on observed survival do not represent a causal estimand, that is, a contrast of potential outcomes for the same set of units under different treatment levels [5, 6]. An approach that avoids this problem and that is often considered in this setting uses the principal stratification approach [7], and involves defining a principal effect for the stratum of the population who would survive to the time of outcome assessment regardless of exposure or treatment level. Because it has a causal interpretation, this would be a suitable target estimand (note that Stensrud and colleagues [8] recently proposed conditional separable effects as another suitable estimand).

Although the principal stratification approach, which is based on the more general potential outcomes framework for causal inference [5, 9], has led to a rich literature on truncation by death, that includes discussions on identification, methods for estimation and sensitivity analyses of principal effects [6, 10-14], a description of truncation by death under other causal frameworks might be valuable [15-18]. In particular, descriptions of causal processes under the sufficient cause framework may be useful [19]. This causal framework is based on the presence or absence of causal mechanisms, which are sufficient causes that once completed invariably lead to the outcome, and its components. This framework might not only provide information on the types of sufficient causes captured by different estimands used in the presence of truncation by death, but also lead to a more in-depth understanding of the potential biases in this context. Indeed, in many areas of epidemiologic research, the framing of questions using the sufficient cause framework has led to better characterisation of key concepts [15-17, 20-28].

In the following section, we briefly summarize the principal stratification approach for truncation by death. We then use the sufficient cause framework to present this problem in the realistic scenario where a common cause of both survival and the outcome is present. It is important to clarify that our goal in providing an account of truncation by death using sufficient causes is not to describe new estimands, identification or estimation approaches for this problem; rather, our objective is to gain better understanding of this problem by considering individual causal mechanisms and their relations to survival and the outcome of interest.

*The principal stratification approach to truncation by death*

Consider a randomized trial recruiting patients with a serious medical condition in which the outcome, $Y$, corresponds to a dichotomized version of a quality of life score one year after trial enrolment (1 = quality of life score above a prespecified threshold, 0 = quality of life score below a prespecified threshold). The treatment is denoted by $X$ (1 = patient is treated, 0 = patient is not treated); survival status one year after enrolment is indicated by the variable $S$ (1 = patient is alive at the time of outcome assessment, 0 = patient died during the follow-up). An essential feature of the problem of truncation by death is that $Y$ is undefined for individuals with $S = 0$. This setting is represented in the directed acyclic graph in **Figure 1** (Panel A), where the variable $U$ denotes common causes of survival and the outcome; here, for concreteness, we let $U$ correspond to a binary variable that denotes age (1 = under 60 years, 0 = 60 years or above), which, we assume, affects both survival and quality of life. In studies affected by truncation by death, it is often assumed that factors that influence both survival and the outcome of interest are present [6, 29].

To define a causal effect that is relevant in this context, we also use the potential outcome variables, $S^x$, which corresponds to the survival status under treatment level $x$, and $Y^x$, the potential value of the outcome variable under treatment level $x$. In **Figure 1** (Panel B), response type variables $S^{\mathrm{T}} = (S^1, S^0)$ and $Y^{\mathrm{T}} = (Y^1, Y^0)$, defined using potential outcomes, are presented [6, 30]. The $S$-related variable can be used to partition the study population into four principal strata: $(S^1, S^0) = (1,1)$, $(S^1, S^0) = (1,0)$, $(S^1, S^0) = (0,1)$, $(S^1, S^0) = (0,0)$. This principal stratification identifies a subgroup of the population for whom an effect of $X$ on $Y$ can be defined, the principal stratum of "always-survivors" $(S^1, S^0) = (1,1)$; unlike the other strata, $Y^x$ is defined under both treatment levels for this stratum. Given the above, the following quantity, often referred to as the survivor average causal effect (SACE) [11, 13], would have a causal interpretation in settings where the outcome is truncated by death:

$$\text{SACE} \triangleq E[Y^1 - Y^0 | (S^1 = 1, S^0 = 1)] \,.$$

Note that the target population of the SACE is the always-survivors, who are not discernible without strong assumptions, as only one of the two potential outcome variables $S^1$ and $S^0$ is observable for a particular individual. Further, the SACE can be also defined for strata of the variable $U$, as shown below:

$$\text{SACE}_u \triangleq E[Y^1 - Y^0 | (S^1 = 1, S^0 = 1), U = u] \,.$$

The $\text{SACE}_u$ may vary across the values of $U$, which is referred to as an effect modification of the SACE [10]. When there is a common cause $U$ of survival and the outcome, an effect modification of the SACE may also occur by another variable that causes survival but does not affect the outcome [10]. Note that the SACE is written as a weighted average of $\text{SACE}_u$, as shown below [10]:

$$\begin{aligned}
\text{SACE} &\triangleq E[Y^1 - Y^0 | (S^1 = 1, S^0 = 1)] \\
&= \sum_u E[Y^1 - Y^0 | (S^1 = 1, S^0 = 1), U = u] \cdot P\big(U = u | (S^1 = 1, S^0 = 1)\big) \\
&= \sum_u \text{SACE}_u \frac{P\big((S^1 = 1, S^0 = 1), U = u\big)}{P\big((S^1 = 1, S^0 = 1)\big)} \\
&= \sum_u \text{SACE}_u \frac{P\big((S^1 = 1, S^0 = 1), U = u\big)}{\sum_{u'} P\big((S^1 = 1, S^0 = 1), U = u'\big)}.
\end{aligned}$$

**Figure 1.** Directed acyclic graphs representing the truncation by death setting. Variables $X$, $S$, and $Y$ denote treatment, survival, and the outcome of interest, respectively; variable $U$ denotes common causes of survival and the outcome. Note, in Panel A, that conditioning on $S = 1$ likely induces collider bias, and that the relationship between $X$ and $U$ is symmetrical; both $X$ and $U$ are common causes of $S$ and $Y$. In Panel B, $S^{\mathrm{T}}$ and $Y^{\mathrm{T}}$, response type variables of $S$ and $Y$, respectively, are also presented. Given that here we consider the SACE, which involves conditioning on $S^{\mathrm{T}} = (1,1)$, Panel B is appropriate [6].

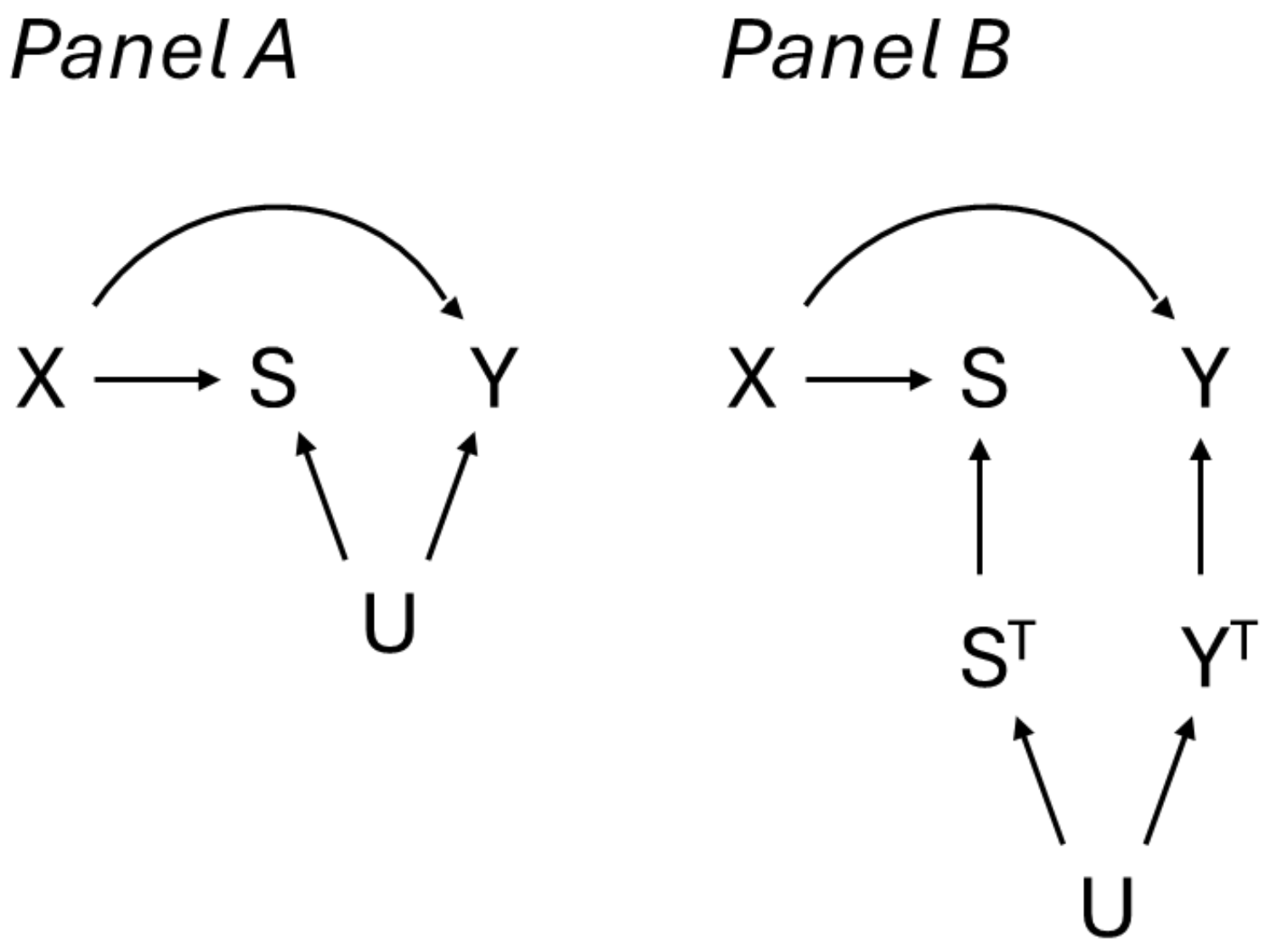

*Truncation by death in the sufficient cause framework*

In the sufficient cause framework, using the definitions of $X$ and $S$ above, and explicitly considering $U$, we can enumerate nine different types of sufficient causes for survival $S$ as follows [31]: $A_1$, $A_2X$, $A_3\overline{X}$, $A_4U$, $A_5\overline{U}$, $A_6UX$, $A_7U\overline{X}$, $A_8\overline{U}X$, and $A_9\overline{UX}$, where $\overline{X}$ and $\overline{U}$ denote the complements of $X$ and $U$, respectively, in the terminology of events, and $A_k$ ($k = 1, \dots, 9$) represents background factors necessary for the completion of the corresponding sufficient cause $k$. The sufficient causes can be interpreted as follows: sufficient cause $A_1$ will lead to survival regardless of the presence or absence of $X$ and $U$; sufficient cause $A_2X$ implies that the presence of $X$ will lead to survival whenever $A_2$ is present (regardless of the presence or absence of $U$); sufficient cause $A_3\overline{X}$ implies that the absence of $X$ will lead to survival whenever $A_3$ is present (regardless of the presence or absence of $U$); sufficient causes $A_4U$ and $A_5\overline{U}$ are analogous to $A_2X$ and $A_3\overline{X}$, but, instead of $X$, are defined based on the presence and absence of $U$, respectively; finally, sufficient causes $A_6UX$, $A_7U\overline{X}$, $A_8\overline{U}X$ and $A_9\overline{UX}$ require either the presence or absence of $X$ and $U$.

Now, if we consider sufficient causes of the outcome "quality of life score above a prespecified threshold" ($Y = 1$), by definition of truncation by death, they all require presence of survival (that is, these sufficient causes require survival as a component cause). Thus, we can enumerate nine different types of sufficient causes for the outcome $Y$ as follows: $B_1S$, $B_2SX$, $B_3S\overline{X}$, $B_4SU$, $B_5S\overline{U}$, $B_6SUX$, $B_7SU\overline{X}$, $B_8S\overline{U}X$, and $B_9S\overline{UX}$, where $B_k$ ($k = 1, \dots, 9$) represents background factors for sufficient cause $k$. Their interpretations are analogous to those in the previous paragraph. Note that we consider two stages of the causal process: the survival stage and the outcome stage; this is analogous to discussions on mediation based on the sufficient cause framework [32], even though survival, in the context of truncation by death, is not a mediator (see Panel A of **Figure 1**).

Here, for ease of discussion and visualization, we assume sufficient cause positive monotonicity for both $X$ and $U$ on $S$ and $Y$ (see **Figure 2**). Put differently, using the definitions of these variables in the previous section, we assume that sufficient causes for survival ($S$) and for quality of life ($Y$) in which either absence of treatment ($\overline{X}$) or age of 60 years or above ($\overline{U}$) are component causes are not present in the population. Thus, only the following four types of sufficient causes for $S$ might be present in the population: $A_1$, $A_2X$, $A_4U$, and $A_6UX$. Similarly, only the following four types of sufficient causes for $Y$ might be present in the population: $B_1S$, $B_2SX$, $B_4SU$, and $B_6SUX$. Consequently, a total of 256 (= $2^4 \times 2^4$) risk status types can be enumerated (**Table S1**), which is defined by possible combinations of susceptibility to sufficient causes [21]. Note that, under the sufficient cause positive monotonicity assumption, the SACE and $\text{SACE}_u$ are non-negative.

Some risk status types only contribute to one of the two terms in $E[Y|S=1, X=1] - E[Y|S=1, X=0]$. Even for the corresponding quantities defined within strata of $U$, that is, $E[Y|S=1, X=1, U=u] - E[Y|S=1, X=0, U=u]$, a subset of the risk status types only contributes to one of the terms. For instance, individuals of risk status types 17 to 32 in **Table S1** are susceptible only to $A_6UX$ for survival. When $X=1$ and $U=1$ (in words, for the subgroup aged less than 60 years who receive treatment), the sufficient cause $A_6UX$ completes; these individuals, by definition, survive and contribute to data on quality of life, and consequently, of $E[Y|S=1, X=1, U=1]$. However, when $X=0$ and $U=1$ (individuals aged less than 60 years who are not treated), no sufficient causes complete for survival; they do not survive, and their quality of life, the outcome $Y$, cannot be defined. Thus, they do not contribute to data of $E[Y|S=1, X=0, U=1]$.

As mentioned above, the target population of the SACE is always-survivors (i.e., $(S^1, S^0) = (1,1)$). Notably, membership to always-survivors is a function of $U$ for some individuals. In other words, an individual of particular risk status types would be in the always-survivors stratum if her or his value of $U$ were 1, but not if this value were 0. For example, let us consider individuals of risk status type 33, who are susceptible only to $A_4U$ for survival (**Table S1**). If they are younger than 60 years ($U=1$), the sufficient cause $A_4U$ completes irrespective of the treatment and they are always-survivors. However, if they are 60 years or older ($U=0$), no sufficient causes complete irrespective of the treatment and they are "never-survivors" (i.e., $(S^1, S^0) = (0,0)$). In fact, under the sufficient cause positive monotonicity assumption, when $U=1$ holds, individuals who are susceptible to $A_1$ or $A_4U$ for survival are always-survivors, and there are 192 (= $3 \times 2^6$) risk status types (**Table 1**). By contrast, when $U=0$ holds, individuals who are susceptible to $A_1$ for survival are always-survivors, and there are 128 (= $2^7$) risk status types (**Table 1**). Therefore, individuals who are not susceptible to $A_1$ but susceptible to $A_4U$ are always-survivors if $U=1$, while they are not if $U=0$; they are either never-survivors (i.e., $(S^1, S^0) = (0,0)$) or survive only under treatment, which is referred to as "protectable" (i.e., $(S^1, S^0) = (1,0)$) (**Table 1**). In the upper part of **Table 2**, we provide expressions of $E[S|X = 1]$, $E[S|X = 0]$, $E[S|X = 1] - E[S|X = 0]$, and $\Pr(S^1 = 1, S^0 = 1)$ in terms of background factor $A_k$ and $U$.

*Conditions under which SACE is null*

Recall that the SACE can be expressed as a weighted average of $\mathrm{SACE}_u$ [10], and since $\mathrm{SACE}_u \geq 0$ holds under the assumption of sufficient cause positive monotonicity, the SACE is null if only and if $\mathrm{SACE}_u$ is null for each level $u$. Therefore, we here describe conditions under which $\mathrm{SACE}_u$ is null – in other words, conditions under which the treatment $X$ has no effect on the outcome $Y$ among the always-survivors with a particular value of $U$. Firstly, among those who are younger than 60 years ($U = 1$), the $\mathrm{SACE}_{u=1}$ is, by definition, null among those with $(A_1 \vee A_4) \wedge \{(B_1 \vee B_4) \vee (\overline{B_2} \wedge \overline{B_6})\}$, which is met by 156 of the 192 risk status types that represent always-survivors under $U = 1$ (that is, those individuals susceptible to $A_1$ or $A_4U$) (**Table 1**). Thus, the $\mathrm{SACE}_{u=1}$ becomes null if and only if there are no individuals with $U = 1$ for whom $(A_1 \vee A_4) \wedge \{(\overline{B_1} \wedge \overline{B_4}) \wedge (B_2 \vee B_6)\}$ holds (i.e., $\Pr\left((A_1 \vee A_4) \wedge \{(\overline{B_1} \wedge \overline{B_4}) \wedge (B_2 \vee B_6)\} \wedge U\right) = 0$) (**Figure 3**). Note that, under this condition, protectable and never-survivors may be present among those with $U = 1$. As an illustration, consider individuals younger than 60 years ($U = 1$) of risk status type 35, who are susceptible only to $A_4U$ for survival $S$ and only to $B_4SU$ for the outcome $Y$ (**Table S1**). As mentioned above, they are always-survivors, and the sufficient cause $B_4SU$ completes to induce the outcome $Y$ irrespective of treatment (i.e., $Y^1 = Y^0 = 1$); the effect of $X$ on $Y$ is null. By contrast, among those who are aged 60 years or older ($U = 0$), the $\mathrm{SACE}_{u=0}$ is null among those with $A_1 \wedge (B_1 \vee \overline{B_2})$, which is met by 96 of the 128 risk status types that represent always-survivors under $U = 0$ (that is, those individuals susceptible to $A_1$) (**Table 1**). Thus, the $\mathrm{SACE}_{u=0}$ becomes null if and only if there are no individuals with $U = 0$ for whom $A_1 \wedge (\overline{B_1} \wedge B_2)$ holds (i.e., $\Pr(A_1 \wedge (\overline{B_1} \wedge B_2) \wedge \overline{U}) = 0$) (**Figure 3**). Note that, under this condition, protectable and never-survivors may be present among those with $U = 0$. For instance, let us consider individuals aged 60 years or older ($U = 0$) of the risk status type 137, who are susceptible only to $A_1$ for survival $S$ and only to $B_1S$ for the outcome $Y$ (**Table S1**). They are always-survivors, and the sufficient cause $B_1S$ completes irrespective of treatment (i.e., $Y^1 = Y^0 = 1$); the effect of $X$ on $Y$ is null among them. To summarize, the conditions given for $\text{SACE}_u$, jointly, correspond to the condition for the SACE to be null (i.e., $\Pr\left((A_1 \vee A_4) \wedge \{(\overline{B_1} \wedge \overline{B_4}) \wedge (B_2 \vee B_6)\} \wedge U\right) = 0$ and $\Pr\left(A_1 \wedge (\overline{B_1} \wedge B_2) \wedge \overline{U}\right) = 0$) (**Figure 3**). In the middle part of **Table 2**, we provide expressions of some estimands including the SACE in terms of background factors $A_k$ and $B_k$ as well as $U$.

Note however that, even if the conditions described above for the $\text{SACE}_u$ is null hold, the crude estimand $E[Y|S = 1, X = 1, U = u] - E[Y|S = 1, X = 0, U = u]$ may not be null. First, let us consider individuals younger than 60 years ( $U = 1$ ). If $\Pr\left((A_1 \vee A_4) \wedge \{(\overline{B_1} \wedge \overline{B_4}) \wedge (B_2 \vee B_6)\} \wedge U\right) = 0$ holds, which means $\text{SACE}_{u=1}$ is null, $E[Y|S = 1, X = 1, U = 1] - E[Y|S = 1, X = 0, U = 1]$ can be expressed as shown below:

$$
\begin{aligned}
&E[Y|S = 1, X = 1, U = 1] - E[Y|S = 1, X = 0, U = 1] \\
&= \Pr((B_1 \vee B_2 \vee B_4 \vee B_6)|(A_1 \vee A_2 \vee A_4 \vee A_6) \wedge U) - \Pr((B_1 \vee B_4)|(A_1 \vee A_4) \wedge U) \\
&= \Pr((B_1 \vee B_4)|(A_1 \vee A_2 \vee A_4 \vee A_6) \wedge U) - \Pr((B_1 \vee B_4)|(A_1 \vee A_4) \wedge U),
\end{aligned}
$$

where the first equality follows from the definitions of the background factors and from the fact that we assume randomization of $X$, and the second equality follows from the assumption that among the individuals with $U = 1$ who are susceptible to $A_1$ or $A_4 U$, only those with $(B_1 \vee B_4) \vee (\overline{B_2} \wedge \overline{B_6})$ may be present in the population (**Figure 3**), and the intersection of $(B_1 \vee B_4) \vee (\overline{B_2} \wedge \overline{B_6})$ and $(B_1 \vee B_2 \vee B_4 \vee B_6)$ is $(B_1 \vee B_4)$. Thus, the crude estimand is null if $(B_1 \vee B_4)$ is equally distributed among individuals with $((A_1 \vee A_2 \vee A_4 \vee A_6) \wedge U)$ and those with $((A_1 \vee A_4) \wedge U)$. Next, let us consider individuals aged 60 years or older ($U = 0$). If $\Pr\left(A_1 \wedge (\overline{B_1} \wedge B_2) \wedge \overline{U}\right) = 0$ holds, which means $\text{SACE}_{u=0}$ is null, $E[Y|S = 1, X = 1, U = 0] - E[Y|S = 1, X = 0, U = 0]$ can be expressed as shown below:

$$
\begin{aligned}
&E[Y|S = 1, X = 1, U = 0] - E[Y|S = 1, X = 0, U = 0] \\
&= \Pr\left((B_1 \vee B_2)|(A_1 \vee A_2) \wedge \overline{U}\right) - \Pr\left(B_1|A_1 \wedge \overline{U}\right) \\
&= \Pr\left(B_1|(A_1 \vee A_2) \wedge \overline{U}\right) - \Pr\left(B_1|A_1 \wedge \overline{U}\right),
\end{aligned}
$$

where the first equality follows from the definitions of the background factors and from the fact that we assume randomization of $X$, and the second equality follows from the assumption that among the individuals with $U = 0$ who are susceptible to $A_1$, only those with $(B_1 \vee \overline{B_2})$ may be present in the population (**Figure 3**), and the intersection of $(B_1 \vee \overline{B_2})$ and $(B_1 \vee B_2)$ is $B_1$. Thus, $E[Y|S=1, X=1, U=0] - E[Y|S=1, X=0, U=0]$ is null if $B_1$ is equally distributed among individuals with $((A_1 \vee A_2) \wedge \overline{U})$ and $(A_1 \wedge \overline{U})$. In the bottom part of **Table 2**, we provide expressions of some estimands including $E[Y|S=1, X=1] - E[Y|S=1, X=0]$ in terms of background factors $A_k$ and $B_k$ as well as $U$.

**Figure 2.** Sufficient causes in the context of truncation by death with a common cause of survival and the outcome. We consider two stages: the survival stage and the outcome stage. The exposure, the survival variable, and the common cause of survival and the outcome are represented by $X$, $S$, and $U$, respectively. We assume sufficient cause positive monotonicity for both $X$ and $U$; note that the relationship between $X$ and $U$ is symmetrical.

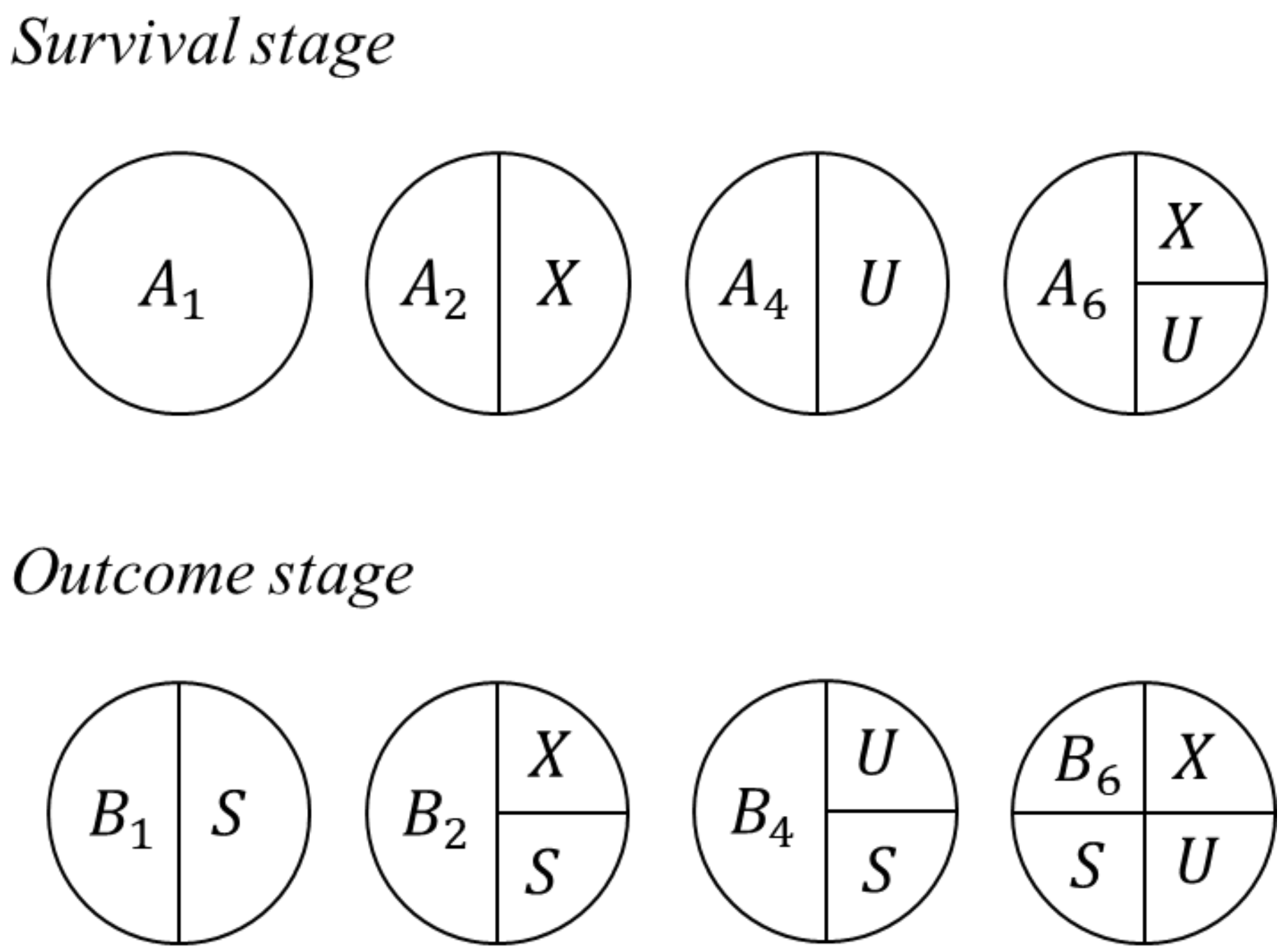

**Figure 3.** Classification of individuals in terms of background factors and $U$ under the sufficient cause positive monotonicity assumption

<table>
<tr><td colspan="2">Subpopulation with $U = 1$</td><td colspan="2">Subpopulation with $U = 0$</td></tr>
<tr><td>Always-survivors ($(S^1, S^0) = (1,1)$) among whom $\text{SACE}_{u=1}$ is null<br><br>$(A_1 \vee A_4) \wedge \{(B_1 \vee B_4) \vee (\overline{B_2} \wedge \overline{B_6})\} \wedge U$</td><td>Always-survivors ($(S^1, S^0) = (1,1)$) among whom $\text{SACE}_{u=1}$ is positive<br><br>$(A_1 \vee A_4) \wedge \{(\overline{B_1} \wedge \overline{B_4}) \wedge (B_2 \vee B_6)\} \wedge U$</td><td>Always-survivors ($(S^1, S^0) = (1,1)$) among whom $\text{SACE}_{u=0}$ is null<br><br>$A_1 \wedge (B_1 \vee \overline{B_2}) \wedge \overline{U}$</td><td>Always-survivors ($(S^1, S^0) = (1,1)$) among whom $\text{SACE}_{u=0}$ is positive<br><br>$A_1 \wedge (\overline{B_1} \wedge B_2) \wedge \overline{U}$</td></tr>
<tr><td colspan="2">Protectable ($(S^1, S^0) = (1,0)$)<br><br>$(\overline{A_1} \wedge \overline{A_4}) \wedge (A_2 \vee A_6) \wedge U$</td><td colspan="2">Protectable ($(S^1, S^0) = (1,0)$)<br><br>$\overline{A_1} \wedge A_2 \wedge \overline{U}$</td></tr>
<tr><td colspan="2">Never-survivors ($(S^1, S^0) = (0,0)$)<br><br>$(\overline{A_1} \wedge \overline{A_4}) \wedge (\overline{A_2} \wedge \overline{A_6}) \wedge U$</td><td colspan="2">Never-survivors ($(S^1, S^0) = (0,0)$)<br><br>$\overline{A_1} \wedge \overline{A_2} \wedge \overline{U}$</td></tr>
</table>

**Table 1.** Descriptions of always-survivors and their risk status types. In the column *Possible risk status types*, the numbers refer to risk status types listed in **Table S1**.

| **Descriptions** | **Expressions in terms of background factors** | **The number of possible risk status types** | **Possible risk status types** |
|---|---|---|---|
| Always-survivors if $U = 1$[a] | $A_1 \vee A_4$ | 192 (= 3 × $2^6$) | 33–64, 97–256 |
| Always-survivors among whom $\text{SACE}_{u=1}$ is null | $(A_1 \vee A_4) \wedge \{(B_1 \vee B_4) \vee (\overline{B_2} \wedge \overline{B_6})\}$ | 156 (= 3 × $2^6 - 3^2 \times 2^2$) | 33, 35, 36, 39–49, 51, 52, 55–64, 97, 99, 100, 103–113, 115, 116, 119–129, 131, 132, 135–145, 147, 148, 151–161, 163, 164, 167–177, 179, 180, 183–193, 195, 196, 199–209, 211, 212, 215–225, 227, 228, 231–241, 243, 244, 247–256 |
| Always-survivors if $U = 0$ | $A_1$ | 128 (= $2^7$) | 129–256 |
| Always-survivors among whom $\text{SACE}_{u=0}$ is null | $A_1 \wedge (B_1 \vee \overline{B_2})$ | 96 (= 3 × $2^5$) | 129–132, 137–148, 153–164, 169–180, 185–196, 201–212, 217–228, 233–244, 249–256 |
| Always-survivors if $U = 1$ but not if $U = 0$ | $\overline{A_1} \wedge A_4$ | 64 (= $2^6$) | 33–64, 97–128 |
| Always-survivors if $U = 1$ but never-survivors if $U = 0$ | $\overline{A_1} \wedge \overline{A_2} \wedge A_4$ | 32 (= $2^5$) | 33–64 |
| Always-survivors if $U = 1$ but protectable if $U = 0$ | $\overline{A_1} \wedge A_2 \wedge A_4$ | 32 (= $2^5$) | 97–128 |

[a]Note that among the subpopulation with $U = 1$, always-survivors (i.e., risk status types 33–64, 97–256) are decomposed to the always-survivors irrespective of $U$ (i.e., risk status types 129–256), never-survivors if $U = 0$ (i.e., risk status types 33–64), and protectable if $U = 0$ (i.e., risk status types 97–128).

**Table 2**. Estimands for truncation by death with a common cause of survival and the outcome. Here, to simplify expressions, we use $\overline{U}$ to denote $U = 0$, and $U$ to denote $U = 1$. Note also that, as we consider the setting of a randomized trial (see section *The principal stratification approach to truncation by death*), the joint distribution of the different background factors and $U$ is the same for both exposed and unexposed individuals; that is, $(A_1, A_2, A_4, A_6, B_1, B_2, B_4, B_6, U) \perp\!\!\!\perp X$. For this reason, we omit the corresponding conjunctions with $X$ and $\overline{X}$ in all the formulas below. Relatedly, note that the distribution of the background factors $A_k$ and $B_k$ might differ by levels of $U$. Note that the expressions below follow from the definitions of the background factors of sufficient causes, and the formula for conditional probabilities; this is evident in the $u$-level specific expressions.

| **Estimands and its terms** | **Expressions in terms of background factors and *U*** |
|---|---|
| $E[S \mid X = 1]$ | $\Pr((A_1 \vee A_2) \wedge \overline{U}) + \Pr((A_1 \vee A_2 \vee A_4 \vee A_6) \wedge U)$ |
| $E[S \mid X = 0]$ | $\Pr(A_1 \wedge \overline{U}) + \Pr((A_1 \vee A_4) \wedge U)$ |
| $E[S \mid X = 1] - E[S \mid X = 0]$ | $\{\Pr((A_1 \vee A_2) \wedge \overline{U}) - \Pr(A_1 \wedge \overline{U})\} + \{\Pr((A_1 \vee A_2 \vee A_4 \vee A_6) \wedge U) - \Pr((A_1 \vee A_4) \wedge U)\}$ |
| $\Pr(S^1 = 1, S^0 = 1)$ | $\Pr(A_1 \wedge \overline{U}) + \Pr((A_1 \vee A_4) \wedge U)$ |
| $E[Y^1 \mid S^1 = 1, S^0 = 1]$ | $\dfrac{\Pr(A_1 \wedge (B_1 \vee B_2) \wedge \overline{U}) + \Pr((A_1 \vee A_4) \wedge (B_1 \vee B_2 \vee B_4 \vee B_6) \wedge U)}{\Pr(A_1 \wedge \overline{U}) + \Pr((A_1 \vee A_4) \wedge U)}$ |
| $E[Y^0 \mid S^1 = 1, S^0 = 1]$ | $\dfrac{\Pr(A_1 \wedge B1 \wedge \overline{U}) + \Pr((A_1 \vee A_4) \wedge (B_1 \vee B_4) \wedge U)}{\Pr(A_1 \wedge \overline{U}) + \Pr((A_1 \vee A_4) \wedge U)}$ |
| $E[Y^1 - Y^0 \mid S^1 = 1, S^0 = 1]$ | $\dfrac{(\Pr(A_1 \wedge (B_1 \vee B_2) \wedge \overline{U}) + \Pr((A_1 \vee A_4) \wedge (B_1 \vee B_2 \vee B_4 \vee B_6) \wedge U)) - (\Pr(A_1 \wedge B_1 \wedge \overline{U}) + \Pr((A_1 \vee A_4) \wedge (B_1 \vee B_4) \wedge U))}{\Pr(A_1 \wedge \overline{U}) + \Pr((A_1 \vee A_4) \wedge U)}$ |
| $E[Y^1 - Y^0 \mid S^1 = 1, S^0 = 1, U = 1]$ | $\dfrac{\Pr((A_1 \vee A_4) \wedge (B_1 \vee B_2 \vee B_4 \vee B_6) \wedge U)) - \Pr((A_1 \vee A_4) \wedge (B_1 \vee B_4) \wedge U)}{\Pr((A_1 \vee A_4) \wedge U)}$ |
| $E[Y^1 - Y^0 \mid S^1 = 1, S^0 = 1, U = 0]$ | $\dfrac{\Pr(A_1 \wedge (B_1 \vee B_2) \wedge \overline{U}) - (\Pr(A_1 \wedge B_1 \wedge \overline{U}))}{\Pr(A_1 \wedge \overline{U})}$ |
| $E[Y \mid S = 1, X = 1]$ | $\dfrac{\Pr((A_1 \vee A_2) \wedge (B_1 \vee B_2) \wedge \overline{U}) + \Pr((A_1 \vee A_2 \vee A_4 \vee A_6) \wedge (B_1 \vee B_2 \vee B_4 \vee B_6) \wedge U)}{\Pr((A_1 \vee A_2) \wedge \overline{U}) + \Pr((A_1 \vee A_2 \vee A_4 \vee A_6) \wedge U)}$ |
| $E[Y\|S=1,X=0]$ | $\frac{\Pr\left(A_1 \wedge B_1 \wedge \overline{U}\right) + \Pr\left((A_1 \vee A_4) \wedge (B_1 \vee B_4) \wedge U\right)}{\Pr\left(A_1 \wedge \overline{U}\right) + \Pr\left((A_1 \vee A_4) \wedge U\right)}$ |
| $E[Y\|S=1,X=1] - E[Y\|S=1,X=0]$ | $\frac{\Pr\left((A_1 \vee A_2) \wedge (B_1 \vee B_2) \wedge \overline{U}\right) + \Pr\left((A_1 \vee A_2 \vee A_4 \vee A_6) \wedge (B_1 \vee B_2 \vee B_4 \vee B_6) \wedge U\right)}{\Pr\left((A_1 \vee A_2) \wedge \overline{U}\right) + \Pr\left((A_1 \vee A_2 \vee A_4 \vee A_6) \wedge U\right)} - \frac{\Pr\left(A_1 \wedge B_1 \wedge \overline{U}\right) + \Pr\left((A_1 \vee A_4) \wedge (B_1 \vee B_4) \wedge U\right)}{\Pr\left(A_1 \wedge \overline{U}\right) + \Pr\left((A_1 \vee A_4) \wedge U\right)}$ |
| $E[Y\|S=1,X=1,U=1] - E[Y\|S=1,X=0,U=1]$ | $\frac{\Pr\left((A_1 \vee A_2 \vee A_4 \vee A_6) \wedge (B_1 \vee B_2 \vee B_4 \vee B_6) \wedge U\right)}{\Pr\left((A_1 \vee A_2 \vee A_4 \vee A_6) \wedge U\right)} - \frac{\Pr\left((A_1 \vee A_4) \wedge (B_1 \vee B_4) \wedge U\right)}{\Pr\left((A_1 \vee A_4) \wedge U\right)}$ |
| $E[Y\|S=1,X=1,U=0] - E[Y\|S=1,X=0,U=0]$ | $\frac{\Pr\left((A_1 \vee A_2) \wedge (B_1 \vee B_2) \wedge \overline{U}\right)}{\Pr\left((A_1 \vee A_2) \wedge \overline{U}\right)} - \frac{\Pr\left(A_1 \wedge B_1 \wedge \overline{U}\right)}{\Pr\left(A_1 \wedge \overline{U}\right)}$ |

*Discussion*

The description of truncation by death under the potential outcomes framework, and specifically under principal stratification, represented an important methodological advance and led to the development of analytical approaches that can be used in this context [11-14]. The sufficient cause framework, in its turn, breaks down causal processes into a finer level of causal mechanisms, that can then be interpreted as determining potential outcomes. As shown in the previous sections, the use of this latter framework to study truncation by death allows: (i) a different interpretation of the frequently-reported estimand that conditions on observed survival, whose non-causal nature can also be understood in terms of risk status types that only contribute data to one of the two terms in the comparison; (ii) a detailed characterization of the distributional properties of sufficient causes that are captured by the SACE (see **Tables 1** and **2**); (iii) establishing conditions defined in terms of sufficient causes under which the causal estimand SACE is equal to zero. Further, the derivation of the expressions for the SACE in terms the background factors could be viewed as an extension of the previous work on the relation between the magnitude of standard causal effects (that is, causal effects that do not condition on joint potential outcomes) and frequencies of background factors [24, 33]. Given this, we believe that the sufficient cause framework, which has been used for decades to study key causal concepts in epidemiology [18, 20-28, 33-35], should also be considered in settings with post-treatment-variable (here, survival)-related problems.

The rationale for focusing the results presented in this paper on $\text{SACE}_u$ is twofold: first, as mentioned above, membership to the always-survivors principal stratum might depend on the value of $U$; second, for the outcome stage, some sufficient causes are dependent on $U$, indicating that these conditions are $u$-level specific. Note that the SACE defined as a risk difference can be expressed as a weighted average of $\text{SACE}_u$ and is a directly collapsible effect measure [10].

Our discussion above also showed that when we consider the two processes of survival and outcome, even if the conditions under which the SACE is null hold, $E[Y|S=1, X=1] - E[Y|S=1, X=0]$ may not be null, even within strata of $U$. This occurs because risk status types that correspond to non-null individual-level effect of $X$ on $S$ will only contribute data to one of these two terms that condition on observed survival, which clearly illustrates the non-causal nature of this estimand.

Another insight from this analysis is that, when considering the presence of possible mechanisms, that is, of sufficient causes, defined not only in terms of the exposure of interest, but also in terms of another cause of survival (here, the variable $U$), for some risk status types, whether an individual is a member of the always-survivors stratum depends on the value of $U$. Put differently, given a population-level distribution of risk status types, the relative frequencies of levels of $U$ might influence the size of the always-survivors stratum in the population, and hence the public health relevance of the SACE. This detailed understanding of processes that contribute to survival, and consequently to principal stratum membership, might encourage descriptions of other questions studied under principal stratification [36] in the sufficient cause framework. For instance, treatment adherence (or compliance), which differs from truncation by death in that the definability of the outcome is not dependent on the post-treatment variable, could also be studied under the sufficient cause framework, possibly with explicit consideration of factors that influence adherence and the outcome. We note, however, that a potential limitation of the application of the sufficient cause framework in truncation by death settings is that this causal framework is more naturally applicable to binary outcome variables, while in studies affected by truncation by death, the outcome of interest is sometimes quantitative (e.g. a quality of life score). Here, it is worth mentioning that one possible extension of this work would involve a formulation of the sufficient cause model using hazard and/or cumulative rates [37] of death.

Taken together, our results suggest that explicit consideration of sufficient causes, even though detailed information on their components is usually not available in real-world data, allows epidemiologists to better understand the mechanisms underlying this problem and the principal stratification approach to define effects in the presence of truncation by death.

*Appendix*

*Derivations used for construction of* ***Table 2***

**Table S1**

*Derivations used for construction of* ***Table 2***

As mentioned in the main text, the SACE is written as a weighted average of SACE$_u$. This can be written using the notation in **Table 2**, as below:

$$E[Y^1 - Y^0|(S^1 = 1, S^0 = 1), U = 1]\frac{P\big((S^1 = 1, S^0 = 1), U = 1\big)}{P\big((S^1 = 1, S^0 = 1)\big)} + E[Y^1 - Y^0|(S^1 = 1, S^0 = 1), U = 0]\frac{P\big((S^1 = 1, S^0 = 1), U = 0\big)}{P\big((S^1 = 1, S^0 = 1)\big)}$$

$$= \frac{\Pr\big((A_1 \vee A_4) \wedge (B_1 \vee B_2 \vee B_4 \vee B_6) \wedge U)\big) - \Pr\big((A_1 \vee A_4) \wedge (B_1 \vee B_4) \wedge U\big)}{\Pr\big((A_1 \vee A_4) \wedge U\big)} \cdot \frac{\Pr\big((A_1 \vee A_4) \wedge U\big)}{\Pr\big(A_1 \wedge \overline{U}\big) + \Pr\big((A_1 \vee A_4) \wedge U\big)}$$

$$+ \frac{\Pr\big(A_1 \wedge (B_1 \vee B_2) \wedge \overline{U}\big) - \big(\Pr\big(A_1 \wedge B_1 \wedge \overline{U}\big)\big)}{\Pr\big(A_1 \wedge \overline{U}\big)} \cdot \frac{\Pr\big(A_1 \wedge \overline{U}\big)}{\Pr\big(A_1 \wedge \overline{U}\big) + \Pr\big((A_1 \vee A_4) \wedge U\big)}$$

$$= \frac{\big(\Pr\big(A_1 \wedge (B_1 \vee B_2) \wedge \overline{U}\big) + \Pr\big((A_1 \vee A_4) \wedge (B_1 \vee B_2 \vee B_4 \vee B_6) \wedge U\big)\big) - \big(\Pr\big(A_1 \wedge B_1 \wedge \overline{U}\big) + \Pr\big((A_1 \vee A_4) \wedge (B_1 \vee B_4) \wedge U\big)\big)}{\Pr\big(A_1 \wedge \overline{U}\big) + \Pr\big((A_1 \vee A_4) \wedge U\big)}$$

$$= E[Y^1 - Y^0|S^1 = 1, S^0 = 1].$$

**Table S1**. Risk status types based on sufficient causes for the truncation by death setting with a common cause of survival and the outcome. Note that in the table, $(S^1S^0) = (1,1)$ is represented by $(AS)$, to indicate the “always survivors” stratum.

| **Risk types** | $A1$ | $A2$ | $A4$ | $A6$ | $B1$ | $B2$ | $B4$ | $B6$ | S\|X=1U=1 | S\|X=0U=1 | S\|X=1U=0 | S\|X=0 U=0 | Y\|S=1 X=1 U=1 | Y\|S=1 X=0 U=1 | Y\|S=1 X=1 U=0 | Y\|S=1 X=0 U=0 | $Y\|(AS)$, $X=1\ U=1$ | $Y\|(AS)$, $X=0\ U=1$ | $Y\|(AS)$, $X=1\ U=0$ | $Y\|(AS)$, $X=0\ U=0$ |
|---|---|---|---|---|---|---|---|---|---|---|---|---|---|---|---|---|---|---|---|---|
| 1 | 0 | 0 | 0 | 0 | 0 | 0 | 0 | 0 | 0 | 0 | 0 | 0 | | | | | | | | |
| 2 | 0 | 0 | 0 | 0 | 0 | 0 | 0 | 1 | 0 | 0 | 0 | 0 | | | | | | | | |
| 3 | 0 | 0 | 0 | 0 | 0 | 0 | 1 | 0 | 0 | 0 | 0 | 0 | | | | | | | | |
| 4 | 0 | 0 | 0 | 0 | 0 | 0 | 1 | 1 | 0 | 0 | 0 | 0 | | | | | | | | |
| 5 | 0 | 0 | 0 | 0 | 0 | 1 | 0 | 0 | 0 | 0 | 0 | 0 | | | | | | | | |
| 6 | 0 | 0 | 0 | 0 | 0 | 1 | 0 | 1 | 0 | 0 | 0 | 0 | | | | | | | | |
| 7 | 0 | 0 | 0 | 0 | 0 | 1 | 1 | 0 | 0 | 0 | 0 | 0 | | | | | | | | |
| 8 | 0 | 0 | 0 | 0 | 0 | 1 | 1 | 1 | 0 | 0 | 0 | 0 | | | | | | | | |
| 9 | 0 | 0 | 0 | 0 | 1 | 0 | 0 | 0 | 0 | 0 | 0 | 0 | | | | | | | | |
| 10 | 0 | 0 | 0 | 0 | 1 | 0 | 0 | 1 | 0 | 0 | 0 | 0 | | | | | | | | |
| 11 | 0 | 0 | 0 | 0 | 1 | 0 | 1 | 0 | 0 | 0 | 0 | 0 | | | | | | | | |
| 12 | 0 | 0 | 0 | 0 | 1 | 0 | 1 | 1 | 0 | 0 | 0 | 0 | | | | | | | | |
| 13 | 0 | 0 | 0 | 0 | 1 | 1 | 0 | 0 | 0 | 0 | 0 | 0 | | | | | | | | |
| 14 | 0 | 0 | 0 | 0 | 1 | 1 | 0 | 1 | 0 | 0 | 0 | 0 | | | | | | | | |
| 15 | 0 | 0 | 0 | 0 | 1 | 1 | 1 | 0 | 0 | 0 | 0 | 0 | | | | | | | | |
| 16 | 0 | 0 | 0 | 0 | 1 | 1 | 1 | 1 | 0 | 0 | 0 | 0 | | | | | | | | |
| 17 | 0 | 0 | 0 | 1 | 0 | 0 | 0 | 0 | 1 | 0 | 0 | 0 | 0 | | | | | | | |
| 18 | 0 | 0 | 0 | 1 | 0 | 0 | 0 | 1 | 1 | 0 | 0 | 0 | 1 | | | | | | | |
| 19 | 0 | 0 | 0 | 1 | 0 | 0 | 1 | 0 | 1 | 0 | 0 | 0 | 1 | | | | | | | |
| 20 | 0 | 0 | 0 | 1 | 0 | 0 | 1 | 1 | 1 | 0 | 0 | 0 | 1 | | | | | | | |
| 21 | 0 | 0 | 0 | 1 | 0 | 1 | 0 | 0 | 1 | 0 | 0 | 0 | 1 | | | | | | | |
| 22 | 0 | 0 | 0 | 1 | 0 | 1 | 0 | 1 | 1 | 0 | 0 | 0 | 1 | | | | | | | |
| 23 | 0 | 0 | 0 | 1 | 0 | 1 | 1 | 0 | 1 | 0 | 0 | 0 | 1 | | | | | | | |
| 24 | 0 | 0 | 0 | 1 | 0 | 1 | 1 | 1 | 1 | 0 | 0 | 0 | 1 | | | | | | | |
| 25 | 0 | 0 | 0 | 1 | 1 | 0 | 0 | 0 | 1 | 0 | 0 | 0 | 1 | | | | | | | |
| 26 | 0 | 0 | 0 | 1 | 1 | 0 | 0 | 1 | 1 | 0 | 0 | 0 | 1 | | | | | | | |
| 27 | 0 | 0 | 0 | 1 | 1 | 0 | 1 | 0 | 1 | 0 | 0 | 0 | 1 | | | | | | | |
| 28 | 0 | 0 | 0 | 1 | 1 | 0 | 1 | 1 | 1 | 0 | 0 | 0 | 1 | | | | | | | |
| 29 | 0 | 0 | 0 | 1 | 1 | 1 | 0 | 0 | 1 | 0 | 0 | 0 | 1 | | | | | | | |
| 30 | 0 | 0 | 0 | 1 | 1 | 1 | 0 | 1 | 1 | 0 | 0 | 0 | 1 | | | | | | | |
| 31 | 0 | 0 | 0 | 1 | 1 | 1 | 1 | 0 | 1 | 0 | 0 | 0 | 1 | | | | | | | |
| 32 | 0 | 0 | 0 | 1 | 1 | 1 | 1 | 1 | 1 | 0 | 0 | 0 | 1 | | | | | | | |
| 33 | 0 | 0 | 1 | 0 | 0 | 0 | 0 | 0 | 1 | 1 | 0 | 0 | 0 | 0 | | | 0 | 0 | | |
| 34 | 0 | 0 | 1 | 0 | 0 | 0 | 0 | 1 | 1 | 1 | 0 | 0 | 1 | 0 | | | 1 | 0 | | |
| 35 | 0 | 0 | 1 | 0 | 0 | 0 | 1 | 0 | 1 | 1 | 0 | 0 | 1 | 1 | | | 1 | 1 | | |
| 36 | 0 | 0 | 1 | 0 | 0 | 0 | 1 | 1 | 1 | 1 | 0 | 0 | 1 | 1 | | | 1 | 1 | | |
| 37 | 0 | 0 | 1 | 0 | 0 | 1 | 0 | 0 | 1 | 1 | 0 | 0 | 1 | 0 | | | 1 | 0 | | |
| 38 | 0 | 0 | 1 | 0 | 0 | 1 | 0 | 1 | 1 | 1 | 0 | 0 | 1 | 0 | | | 1 | 0 | | |

| | | | | | | | | | | | | | | | | | |
|---|---|---|---|---|---|---|---|---|---|---|---|---|---|---|---|---|---|
| 39 | 0 | 0 | 1 | 0 | 0 | 1 | 1 | 0 | 1 | 1 | 0 | 0 | 1 | 1 | | 1 | 1 |
| 40 | 0 | 0 | 1 | 0 | 0 | 1 | 1 | 1 | 1 | 1 | 0 | 0 | 1 | 1 | | 1 | 1 |
| 41 | 0 | 0 | 1 | 0 | 1 | 0 | 0 | 0 | 1 | 1 | 0 | 0 | 1 | 1 | | 1 | 1 |
| 42 | 0 | 0 | 1 | 0 | 1 | 0 | 0 | 1 | 1 | 1 | 0 | 0 | 1 | 1 | | 1 | 1 |
| 43 | 0 | 0 | 1 | 0 | 1 | 0 | 1 | 0 | 1 | 1 | 0 | 0 | 1 | 1 | | 1 | 1 |
| 44 | 0 | 0 | 1 | 0 | 1 | 0 | 1 | 1 | 1 | 1 | 0 | 0 | 1 | 1 | | 1 | 1 |
| 45 | 0 | 0 | 1 | 0 | 1 | 1 | 0 | 0 | 1 | 1 | 0 | 0 | 1 | 1 | | 1 | 1 |
| 46 | 0 | 0 | 1 | 0 | 1 | 1 | 0 | 1 | 1 | 1 | 0 | 0 | 1 | 1 | | 1 | 1 |
| 47 | 0 | 0 | 1 | 0 | 1 | 1 | 1 | 0 | 1 | 1 | 0 | 0 | 1 | 1 | | 1 | 1 |
| 48 | 0 | 0 | 1 | 0 | 1 | 1 | 1 | 1 | 1 | 1 | 0 | 0 | 1 | 1 | | 1 | 1 |
| 49 | 0 | 0 | 1 | 1 | 0 | 0 | 0 | 0 | 1 | 1 | 0 | 0 | 0 | 0 | | 0 | 0 |
| 50 | 0 | 0 | 1 | 1 | 0 | 0 | 0 | 1 | 1 | 1 | 0 | 0 | 1 | 0 | | 1 | 0 |
| 51 | 0 | 0 | 1 | 1 | 0 | 0 | 1 | 0 | 1 | 1 | 0 | 0 | 1 | 1 | | 1 | 1 |
| 52 | 0 | 0 | 1 | 1 | 0 | 0 | 1 | 1 | 1 | 1 | 0 | 0 | 1 | 1 | | 1 | 1 |
| 53 | 0 | 0 | 1 | 1 | 0 | 1 | 0 | 0 | 1 | 1 | 0 | 0 | 1 | 0 | | 1 | 0 |
| 54 | 0 | 0 | 1 | 1 | 0 | 1 | 0 | 1 | 1 | 1 | 0 | 0 | 1 | 0 | | 1 | 0 |
| 55 | 0 | 0 | 1 | 1 | 0 | 1 | 1 | 0 | 1 | 1 | 0 | 0 | 1 | 1 | | 1 | 1 |
| 56 | 0 | 0 | 1 | 1 | 0 | 1 | 1 | 1 | 1 | 1 | 0 | 0 | 1 | 1 | | 1 | 1 |
| 57 | 0 | 0 | 1 | 1 | 1 | 0 | 0 | 0 | 1 | 1 | 0 | 0 | 1 | 1 | | 1 | 1 |
| 58 | 0 | 0 | 1 | 1 | 1 | 0 | 0 | 1 | 1 | 1 | 0 | 0 | 1 | 1 | | 1 | 1 |
| 59 | 0 | 0 | 1 | 1 | 1 | 0 | 1 | 0 | 1 | 1 | 0 | 0 | 1 | 1 | | 1 | 1 |
| 60 | 0 | 0 | 1 | 1 | 1 | 0 | 1 | 1 | 1 | 1 | 0 | 0 | 1 | 1 | | 1 | 1 |
| 61 | 0 | 0 | 1 | 1 | 1 | 1 | 0 | 0 | 1 | 1 | 0 | 0 | 1 | 1 | | 1 | 1 |
| 62 | 0 | 0 | 1 | 1 | 1 | 1 | 0 | 1 | 1 | 1 | 0 | 0 | 1 | 1 | | 1 | 1 |
| 63 | 0 | 0 | 1 | 1 | 1 | 1 | 1 | 0 | 1 | 1 | 0 | 0 | 1 | 1 | | 1 | 1 |
| 64 | 0 | 0 | 1 | 1 | 1 | 1 | 1 | 1 | 1 | 1 | 0 | 0 | 1 | 1 | | 1 | 1 |
| 65 | 0 | 1 | 0 | 0 | 0 | 0 | 0 | 0 | 1 | 0 | 1 | 0 | 0 | | 0 | | |
| 66 | 0 | 1 | 0 | 0 | 0 | 0 | 0 | 1 | 1 | 0 | 1 | 0 | 1 | | 0 | | |
| 67 | 0 | 1 | 0 | 0 | 0 | 0 | 1 | 0 | 1 | 0 | 1 | 0 | 1 | | 0 | | |
| 68 | 0 | 1 | 0 | 0 | 0 | 0 | 1 | 1 | 1 | 0 | 1 | 0 | 1 | | 0 | | |
| 69 | 0 | 1 | 0 | 0 | 0 | 1 | 0 | 0 | 1 | 0 | 1 | 0 | 1 | | 1 | | |
| 70 | 0 | 1 | 0 | 0 | 0 | 1 | 0 | 1 | 1 | 0 | 1 | 0 | 1 | | 1 | | |
| 71 | 0 | 1 | 0 | 0 | 0 | 1 | 1 | 0 | 1 | 0 | 1 | 0 | 1 | | 1 | | |
| 72 | 0 | 1 | 0 | 0 | 0 | 1 | 1 | 1 | 1 | 0 | 1 | 0 | 1 | | 1 | | |
| 73 | 0 | 1 | 0 | 0 | 1 | 0 | 0 | 0 | 1 | 0 | 1 | 0 | 1 | | 1 | | |
| 74 | 0 | 1 | 0 | 0 | 1 | 0 | 0 | 1 | 1 | 0 | 1 | 0 | 1 | | 1 | | |
| 75 | 0 | 1 | 0 | 0 | 1 | 0 | 1 | 0 | 1 | 0 | 1 | 0 | 1 | | 1 | | |
| 76 | 0 | 1 | 0 | 0 | 1 | 0 | 1 | 1 | 1 | 0 | 1 | 0 | 1 | | 1 | | |
| 77 | 0 | 1 | 0 | 0 | 1 | 1 | 0 | 0 | 1 | 0 | 1 | 0 | 1 | | 1 | | |
| 78 | 0 | 1 | 0 | 0 | 1 | 1 | 0 | 1 | 1 | 0 | 1 | 0 | 1 | | 1 | | |
| 79 | 0 | 1 | 0 | 0 | 1 | 1 | 1 | 0 | 1 | 0 | 1 | 0 | 1 | | 1 | | |
| 80 | 0 | 1 | 0 | 0 | 1 | 1 | 1 | 1 | 1 | 0 | 1 | 0 | 1 | | 1 | | |
| 81 | 0 | 1 | 0 | 1 | 0 | 0 | 0 | 0 | 1 | 0 | 1 | 0 | 0 | | 0 | | |
| 82 | 0 | 1 | 0 | 1 | 0 | 0 | 0 | 1 | 1 | 0 | 1 | 0 | 1 | | 0 | | |
| 83 | 0 | 1 | 0 | 1 | 0 | 0 | 1 | 0 | 1 | 0 | 1 | 0 | 1 | | 0 | | |
| 84 | 0 | 1 | 0 | 1 | 0 | 0 | 1 | 1 | 1 | 0 | 1 | 0 | 1 | | 0 | | |
| 85 | 0 | 1 | 0 | 1 | 0 | 1 | 0 | 0 | 1 | 0 | 1 | 0 | 1 | | 1 | | |
| 86 | 0 | 1 | 0 | 1 | 0 | 1 | 0 | 1 | 1 | 0 | 1 | 0 | 1 | | 1 | | |
| 87 | 0 | 1 | 0 | 1 | 0 | 1 | 1 | 0 | 1 | 0 | 1 | 0 | 1 | | 1 | | |

| | | | | | | | | | | | | | | | | | | | | |
|---|---|---|---|---|---|---|---|---|---|---|---|---|---|---|---|---|---|---|---|---|
| 88 | 0 | 1 | 0 | 1 | 0 | 1 | 1 | 1 | 1 | 0 | 1 | 0 | 1 | | 1 | | | | | |
| 89 | 0 | 1 | 0 | 1 | 1 | 0 | 0 | 0 | 1 | 0 | 1 | 0 | 1 | | 1 | | | | | |
| 90 | 0 | 1 | 0 | 1 | 1 | 0 | 0 | 1 | 1 | 0 | 1 | 0 | 1 | | 1 | | | | | |
| 91 | 0 | 1 | 0 | 1 | 1 | 0 | 1 | 0 | 1 | 0 | 1 | 0 | 1 | | 1 | | | | | |
| 92 | 0 | 1 | 0 | 1 | 1 | 0 | 1 | 1 | 1 | 0 | 1 | 0 | 1 | | 1 | | | | | |
| 93 | 0 | 1 | 0 | 1 | 1 | 1 | 0 | 0 | 1 | 0 | 1 | 0 | 1 | | 1 | | | | | |
| 94 | 0 | 1 | 0 | 1 | 1 | 1 | 0 | 1 | 1 | 0 | 1 | 0 | 1 | | 1 | | | | | |
| 95 | 0 | 1 | 0 | 1 | 1 | 1 | 1 | 0 | 1 | 0 | 1 | 0 | 1 | | 1 | | | | | |
| 96 | 0 | 1 | 0 | 1 | 1 | 1 | 1 | 1 | 1 | 0 | 1 | 0 | 1 | | 1 | | | | | |
| 97 | 0 | 1 | 1 | 0 | 0 | 0 | 0 | 0 | 1 | 1 | 1 | 0 | 0 | 0 | 0 | | 0 | 0 | | |
| 98 | 0 | 1 | 1 | 0 | 0 | 0 | 0 | 1 | 1 | 1 | 1 | 0 | 1 | 0 | 0 | | 1 | 0 | | |
| 99 | 0 | 1 | 1 | 0 | 0 | 0 | 1 | 0 | 1 | 1 | 1 | 0 | 1 | 1 | 0 | | 1 | 1 | | |
| 100 | 0 | 1 | 1 | 0 | 0 | 0 | 1 | 1 | 1 | 1 | 1 | 0 | 1 | 1 | 0 | | 1 | 1 | | |
| 101 | 0 | 1 | 1 | 0 | 0 | 1 | 0 | 0 | 1 | 1 | 1 | 0 | 1 | 0 | 1 | | 1 | 0 | | |
| 102 | 0 | 1 | 1 | 0 | 0 | 1 | 0 | 1 | 1 | 1 | 1 | 0 | 1 | 0 | 1 | | 1 | 0 | | |
| 103 | 0 | 1 | 1 | 0 | 0 | 1 | 1 | 0 | 1 | 1 | 1 | 0 | 1 | 1 | 1 | | 1 | 1 | | |
| 104 | 0 | 1 | 1 | 0 | 0 | 1 | 1 | 1 | 1 | 1 | 1 | 0 | 1 | 1 | 1 | | 1 | 1 | | |
| 105 | 0 | 1 | 1 | 0 | 1 | 0 | 0 | 0 | 1 | 1 | 1 | 0 | 1 | 1 | 1 | | 1 | 1 | | |
| 106 | 0 | 1 | 1 | 0 | 1 | 0 | 0 | 1 | 1 | 1 | 1 | 0 | 1 | 1 | 1 | | 1 | 1 | | |
| 107 | 0 | 1 | 1 | 0 | 1 | 0 | 1 | 0 | 1 | 1 | 1 | 0 | 1 | 1 | 1 | | 1 | 1 | | |
| 108 | 0 | 1 | 1 | 0 | 1 | 0 | 1 | 1 | 1 | 1 | 1 | 0 | 1 | 1 | 1 | | 1 | 1 | | |
| 109 | 0 | 1 | 1 | 0 | 1 | 1 | 0 | 0 | 1 | 1 | 1 | 0 | 1 | 1 | 1 | | 1 | 1 | | |
| 110 | 0 | 1 | 1 | 0 | 1 | 1 | 0 | 1 | 1 | 1 | 1 | 0 | 1 | 1 | 1 | | 1 | 1 | | |
| 111 | 0 | 1 | 1 | 0 | 1 | 1 | 1 | 0 | 1 | 1 | 1 | 0 | 1 | 1 | 1 | | 1 | 1 | | |
| 112 | 0 | 1 | 1 | 0 | 1 | 1 | 1 | 1 | 1 | 1 | 1 | 0 | 1 | 1 | 1 | | 1 | 1 | | |
| 113 | 0 | 1 | 1 | 1 | 0 | 0 | 0 | 0 | 1 | 1 | 1 | 0 | 0 | 0 | 0 | | 0 | 0 | | |
| 114 | 0 | 1 | 1 | 1 | 0 | 0 | 0 | 1 | 1 | 1 | 1 | 0 | 1 | 0 | 0 | | 1 | 0 | | |
| 115 | 0 | 1 | 1 | 1 | 0 | 0 | 1 | 0 | 1 | 1 | 1 | 0 | 1 | 1 | 0 | | 1 | 1 | | |
| 116 | 0 | 1 | 1 | 1 | 0 | 0 | 1 | 1 | 1 | 1 | 1 | 0 | 1 | 1 | 0 | | 1 | 1 | | |
| 117 | 0 | 1 | 1 | 1 | 0 | 1 | 0 | 0 | 1 | 1 | 1 | 0 | 1 | 0 | 1 | | 1 | 0 | | |
| 118 | 0 | 1 | 1 | 1 | 0 | 1 | 0 | 1 | 1 | 1 | 1 | 0 | 1 | 0 | 1 | | 1 | 0 | | |
| 119 | 0 | 1 | 1 | 1 | 0 | 1 | 1 | 0 | 1 | 1 | 1 | 0 | 1 | 1 | 1 | | 1 | 1 | | |
| 120 | 0 | 1 | 1 | 1 | 0 | 1 | 1 | 1 | 1 | 1 | 1 | 0 | 1 | 1 | 1 | | 1 | 1 | | |
| 121 | 0 | 1 | 1 | 1 | 1 | 0 | 0 | 0 | 1 | 1 | 1 | 0 | 1 | 1 | 1 | | 1 | 1 | | |
| 122 | 0 | 1 | 1 | 1 | 1 | 0 | 0 | 1 | 1 | 1 | 1 | 0 | 1 | 1 | 1 | | 1 | 1 | | |
| 123 | 0 | 1 | 1 | 1 | 1 | 0 | 1 | 0 | 1 | 1 | 1 | 0 | 1 | 1 | 1 | | 1 | 1 | | |
| 124 | 0 | 1 | 1 | 1 | 1 | 0 | 1 | 1 | 1 | 1 | 1 | 0 | 1 | 1 | 1 | | 1 | 1 | | |
| 125 | 0 | 1 | 1 | 1 | 1 | 1 | 0 | 0 | 1 | 1 | 1 | 0 | 1 | 1 | 1 | | 1 | 1 | | |
| 126 | 0 | 1 | 1 | 1 | 1 | 1 | 0 | 1 | 1 | 1 | 1 | 0 | 1 | 1 | 1 | | 1 | 1 | | |
| 127 | 0 | 1 | 1 | 1 | 1 | 1 | 1 | 0 | 1 | 1 | 1 | 0 | 1 | 1 | 1 | | 1 | 1 | | |
| 128 | 0 | 1 | 1 | 1 | 1 | 1 | 1 | 1 | 1 | 1 | 1 | 0 | 1 | 1 | 1 | | 1 | 1 | | |
| 129 | 1 | 0 | 0 | 0 | 0 | 0 | 0 | 0 | 1 | 1 | 1 | 1 | 0 | 0 | 0 | 0 | 0 | 0 | 0 | 0 |
| 130 | 1 | 0 | 0 | 0 | 0 | 0 | 0 | 1 | 1 | 1 | 1 | 1 | 1 | 0 | 0 | 0 | 1 | 0 | 0 | 0 |
| 131 | 1 | 0 | 0 | 0 | 0 | 0 | 1 | 0 | 1 | 1 | 1 | 1 | 1 | 1 | 0 | 0 | 1 | 1 | 0 | 0 |
| 132 | 1 | 0 | 0 | 0 | 0 | 0 | 1 | 1 | 1 | 1 | 1 | 1 | 1 | 1 | 0 | 0 | 1 | 1 | 0 | 0 |
| 133 | 1 | 0 | 0 | 0 | 0 | 1 | 0 | 0 | 1 | 1 | 1 | 1 | 1 | 0 | 1 | 0 | 1 | 0 | 1 | 0 |
| 134 | 1 | 0 | 0 | 0 | 0 | 1 | 0 | 1 | 1 | 1 | 1 | 1 | 1 | 0 | 1 | 0 | 1 | 0 | 1 | 0 |
| 135 | 1 | 0 | 0 | 0 | 0 | 1 | 1 | 0 | 1 | 1 | 1 | 1 | 1 | 1 | 1 | 0 | 1 | 1 | 1 | 0 |
| 136 | 1 | 0 | 0 | 0 | 0 | 1 | 1 | 1 | 1 | 1 | 1 | 1 | 1 | 1 | 1 | 0 | 1 | 1 | 1 | 0 |

| | | | | | | | | | | | | | | | | | | | | | | | | |
|---|---|---|---|---|---|---|---|---|---|---|---|---|---|---|---|---|---|---|---|---|---|---|---|---|
| 137 | 1 | 0 | 0 | 0 | 1 | 0 | 0 | 0 | 1 | 1 | 1 | 1 | 1 | 1 | 1 | 1 | 1 | 1 | 1 | 1 |
| 138 | 1 | 0 | 0 | 0 | 1 | 0 | 0 | 1 | 1 | 1 | 1 | 1 | 1 | 1 | 1 | 1 | 1 | 1 | 1 | 1 |
| 139 | 1 | 0 | 0 | 0 | 1 | 0 | 1 | 0 | 1 | 1 | 1 | 1 | 1 | 1 | 1 | 1 | 1 | 1 | 1 | 1 |
| 140 | 1 | 0 | 0 | 0 | 1 | 0 | 1 | 1 | 1 | 1 | 1 | 1 | 1 | 1 | 1 | 1 | 1 | 1 | 1 | 1 |
| 141 | 1 | 0 | 0 | 0 | 1 | 1 | 0 | 0 | 1 | 1 | 1 | 1 | 1 | 1 | 1 | 1 | 1 | 1 | 1 | 1 |
| 142 | 1 | 0 | 0 | 0 | 1 | 1 | 0 | 1 | 1 | 1 | 1 | 1 | 1 | 1 | 1 | 1 | 1 | 1 | 1 | 1 |
| 143 | 1 | 0 | 0 | 0 | 1 | 1 | 1 | 0 | 1 | 1 | 1 | 1 | 1 | 1 | 1 | 1 | 1 | 1 | 1 | 1 |
| 144 | 1 | 0 | 0 | 0 | 1 | 1 | 1 | 1 | 1 | 1 | 1 | 1 | 1 | 1 | 1 | 1 | 1 | 1 | 1 | 1 |
| 145 | 1 | 0 | 0 | 1 | 0 | 0 | 0 | 0 | 1 | 1 | 1 | 1 | 0 | 0 | 0 | 0 | 0 | 0 | 0 | 0 |
| 146 | 1 | 0 | 0 | 1 | 0 | 0 | 0 | 1 | 1 | 1 | 1 | 1 | 1 | 0 | 0 | 0 | 1 | 0 | 0 | 0 |
| 147 | 1 | 0 | 0 | 1 | 0 | 0 | 1 | 0 | 1 | 1 | 1 | 1 | 1 | 1 | 0 | 0 | 1 | 1 | 0 | 0 |
| 148 | 1 | 0 | 0 | 1 | 0 | 0 | 1 | 1 | 1 | 1 | 1 | 1 | 1 | 1 | 0 | 0 | 1 | 1 | 0 | 0 |
| 149 | 1 | 0 | 0 | 1 | 0 | 1 | 0 | 0 | 1 | 1 | 1 | 1 | 1 | 0 | 1 | 0 | 1 | 0 | 1 | 0 |
| 150 | 1 | 0 | 0 | 1 | 0 | 1 | 0 | 1 | 1 | 1 | 1 | 1 | 1 | 0 | 1 | 0 | 1 | 0 | 1 | 0 |
| 151 | 1 | 0 | 0 | 1 | 0 | 1 | 1 | 0 | 1 | 1 | 1 | 1 | 1 | 1 | 1 | 0 | 1 | 1 | 1 | 0 |
| 152 | 1 | 0 | 0 | 1 | 0 | 1 | 1 | 1 | 1 | 1 | 1 | 1 | 1 | 1 | 1 | 0 | 1 | 1 | 1 | 0 |
| 153 | 1 | 0 | 0 | 1 | 1 | 0 | 0 | 0 | 1 | 1 | 1 | 1 | 1 | 1 | 1 | 1 | 1 | 1 | 1 | 1 |
| 154 | 1 | 0 | 0 | 1 | 1 | 0 | 0 | 1 | 1 | 1 | 1 | 1 | 1 | 1 | 1 | 1 | 1 | 1 | 1 | 1 |
| 155 | 1 | 0 | 0 | 1 | 1 | 0 | 1 | 0 | 1 | 1 | 1 | 1 | 1 | 1 | 1 | 1 | 1 | 1 | 1 | 1 |
| 156 | 1 | 0 | 0 | 1 | 1 | 0 | 1 | 1 | 1 | 1 | 1 | 1 | 1 | 1 | 1 | 1 | 1 | 1 | 1 | 1 |
| 157 | 1 | 0 | 0 | 1 | 1 | 1 | 0 | 0 | 1 | 1 | 1 | 1 | 1 | 1 | 1 | 1 | 1 | 1 | 1 | 1 |
| 158 | 1 | 0 | 0 | 1 | 1 | 1 | 0 | 1 | 1 | 1 | 1 | 1 | 1 | 1 | 1 | 1 | 1 | 1 | 1 | 1 |
| 159 | 1 | 0 | 0 | 1 | 1 | 1 | 1 | 0 | 1 | 1 | 1 | 1 | 1 | 1 | 1 | 1 | 1 | 1 | 1 | 1 |
| 160 | 1 | 0 | 0 | 1 | 1 | 1 | 1 | 1 | 1 | 1 | 1 | 1 | 1 | 1 | 1 | 1 | 1 | 1 | 1 | 1 |
| 161 | 1 | 0 | 1 | 0 | 0 | 0 | 0 | 0 | 1 | 1 | 1 | 1 | 0 | 0 | 0 | 0 | 0 | 0 | 0 | 0 |
| 162 | 1 | 0 | 1 | 0 | 0 | 0 | 0 | 1 | 1 | 1 | 1 | 1 | 1 | 0 | 0 | 0 | 1 | 0 | 0 | 0 |
| 163 | 1 | 0 | 1 | 0 | 0 | 0 | 1 | 0 | 1 | 1 | 1 | 1 | 1 | 1 | 0 | 0 | 1 | 1 | 0 | 0 |
| 164 | 1 | 0 | 1 | 0 | 0 | 0 | 1 | 1 | 1 | 1 | 1 | 1 | 1 | 1 | 0 | 0 | 1 | 1 | 0 | 0 |
| 165 | 1 | 0 | 1 | 0 | 0 | 1 | 0 | 0 | 1 | 1 | 1 | 1 | 1 | 0 | 1 | 0 | 1 | 0 | 1 | 0 |
| 166 | 1 | 0 | 1 | 0 | 0 | 1 | 0 | 1 | 1 | 1 | 1 | 1 | 1 | 0 | 1 | 0 | 1 | 0 | 1 | 0 |
| 167 | 1 | 0 | 1 | 0 | 0 | 1 | 1 | 0 | 1 | 1 | 1 | 1 | 1 | 1 | 1 | 0 | 1 | 1 | 1 | 0 |
| 168 | 1 | 0 | 1 | 0 | 0 | 1 | 1 | 1 | 1 | 1 | 1 | 1 | 1 | 1 | 1 | 0 | 1 | 1 | 1 | 0 |
| 169 | 1 | 0 | 1 | 0 | 1 | 0 | 0 | 0 | 1 | 1 | 1 | 1 | 1 | 1 | 1 | 1 | 1 | 1 | 1 | 1 |
| 170 | 1 | 0 | 1 | 0 | 1 | 0 | 0 | 1 | 1 | 1 | 1 | 1 | 1 | 1 | 1 | 1 | 1 | 1 | 1 | 1 |
| 171 | 1 | 0 | 1 | 0 | 1 | 0 | 1 | 0 | 1 | 1 | 1 | 1 | 1 | 1 | 1 | 1 | 1 | 1 | 1 | 1 |
| 172 | 1 | 0 | 1 | 0 | 1 | 0 | 1 | 1 | 1 | 1 | 1 | 1 | 1 | 1 | 1 | 1 | 1 | 1 | 1 | 1 |
| 173 | 1 | 0 | 1 | 0 | 1 | 1 | 0 | 0 | 1 | 1 | 1 | 1 | 1 | 1 | 1 | 1 | 1 | 1 | 1 | 1 |
| 174 | 1 | 0 | 1 | 0 | 1 | 1 | 0 | 1 | 1 | 1 | 1 | 1 | 1 | 1 | 1 | 1 | 1 | 1 | 1 | 1 |
| 175 | 1 | 0 | 1 | 0 | 1 | 1 | 1 | 0 | 1 | 1 | 1 | 1 | 1 | 1 | 1 | 1 | 1 | 1 | 1 | 1 |
| 176 | 1 | 0 | 1 | 0 | 1 | 1 | 1 | 1 | 1 | 1 | 1 | 1 | 1 | 1 | 1 | 1 | 1 | 1 | 1 | 1 |
| 177 | 1 | 0 | 1 | 1 | 0 | 0 | 0 | 0 | 1 | 1 | 1 | 1 | 0 | 0 | 0 | 0 | 0 | 0 | 0 | 0 |
| 178 | 1 | 0 | 1 | 1 | 0 | 0 | 0 | 1 | 1 | 1 | 1 | 1 | 1 | 0 | 0 | 0 | 1 | 0 | 0 | 0 |
| 179 | 1 | 0 | 1 | 1 | 0 | 0 | 1 | 0 | 1 | 1 | 1 | 1 | 1 | 1 | 0 | 0 | 1 | 1 | 0 | 0 |
| 180 | 1 | 0 | 1 | 1 | 0 | 0 | 1 | 1 | 1 | 1 | 1 | 1 | 1 | 1 | 0 | 0 | 1 | 1 | 0 | 0 |
| 181 | 1 | 0 | 1 | 1 | 0 | 1 | 0 | 0 | 1 | 1 | 1 | 1 | 1 | 0 | 1 | 0 | 1 | 0 | 1 | 0 |
| 182 | 1 | 0 | 1 | 1 | 0 | 1 | 0 | 1 | 1 | 1 | 1 | 1 | 1 | 0 | 1 | 0 | 1 | 0 | 1 | 0 |
| 183 | 1 | 0 | 1 | 1 | 0 | 1 | 1 | 0 | 1 | 1 | 1 | 1 | 1 | 1 | 1 | 0 | 1 | 1 | 1 | 0 |
| 184 | 1 | 0 | 1 | 1 | 0 | 1 | 1 | 1 | 1 | 1 | 1 | 1 | 1 | 1 | 1 | 0 | 1 | 1 | 1 | 0 |
| 185 | 1 | 0 | 1 | 1 | 1 | 0 | 0 | 0 | 1 | 1 | 1 | 1 | 1 | 1 | 1 | 1 | 1 | 1 | 1 | 1 |
| 186 | 1 | 0 | 1 | 1 | 1 | 0 | 0 | 1 | 1 | 1 | 1 | 1 | 1 | 1 | 1 | 1 | 1 | 1 | 1 | 1 |
| 187 | 1 | 0 | 1 | 1 | 1 | 0 | 1 | 0 | 1 | 1 | 1 | 1 | 1 | 1 | 1 | 1 | 1 | 1 | 1 | 1 |
| 188 | 1 | 0 | 1 | 1 | 1 | 0 | 1 | 1 | 1 | 1 | 1 | 1 | 1 | 1 | 1 | 1 | 1 | 1 | 1 | 1 |
| 189 | 1 | 0 | 1 | 1 | 1 | 1 | 0 | 0 | 1 | 1 | 1 | 1 | 1 | 1 | 1 | 1 | 1 | 1 | 1 | 1 |
| 190 | 1 | 0 | 1 | 1 | 1 | 1 | 0 | 1 | 1 | 1 | 1 | 1 | 1 | 1 | 1 | 1 | 1 | 1 | 1 | 1 |
| 191 | 1 | 0 | 1 | 1 | 1 | 1 | 1 | 0 | 1 | 1 | 1 | 1 | 1 | 1 | 1 | 1 | 1 | 1 | 1 | 1 |
| 192 | 1 | 0 | 1 | 1 | 1 | 1 | 1 | 1 | 1 | 1 | 1 | 1 | 1 | 1 | 1 | 1 | 1 | 1 | 1 | 1 |
| 193 | 1 | 1 | 0 | 0 | 0 | 0 | 0 | 0 | 1 | 1 | 1 | 1 | 0 | 0 | 0 | 0 | 0 | 0 | 0 | 0 |
| 194 | 1 | 1 | 0 | 0 | 0 | 0 | 0 | 1 | 1 | 1 | 1 | 1 | 1 | 0 | 0 | 0 | 1 | 0 | 0 | 0 |
| 195 | 1 | 1 | 0 | 0 | 0 | 0 | 1 | 0 | 1 | 1 | 1 | 1 | 1 | 1 | 0 | 0 | 1 | 1 | 0 | 0 |
| 196 | 1 | 1 | 0 | 0 | 0 | 0 | 1 | 1 | 1 | 1 | 1 | 1 | 1 | 1 | 0 | 0 | 1 | 1 | 0 | 0 |
| 197 | 1 | 1 | 0 | 0 | 0 | 1 | 0 | 0 | 1 | 1 | 1 | 1 | 1 | 0 | 1 | 0 | 1 | 0 | 1 | 0 |
| 198 | 1 | 1 | 0 | 0 | 0 | 1 | 0 | 1 | 1 | 1 | 1 | 1 | 1 | 0 | 1 | 0 | 1 | 0 | 1 | 0 |
| 199 | 1 | 1 | 0 | 0 | 0 | 1 | 1 | 0 | 1 | 1 | 1 | 1 | 1 | 1 | 1 | 0 | 1 | 1 | 1 | 0 |
| 200 | 1 | 1 | 0 | 0 | 0 | 1 | 1 | 1 | 1 | 1 | 1 | 1 | 1 | 1 | 1 | 0 | 1 | 1 | 1 | 0 |
| 201 | 1 | 1 | 0 | 0 | 1 | 0 | 0 | 0 | 1 | 1 | 1 | 1 | 1 | 1 | 1 | 1 | 1 | 1 | 1 | 1 |
| 202 | 1 | 1 | 0 | 0 | 1 | 0 | 0 | 1 | 1 | 1 | 1 | 1 | 1 | 1 | 1 | 1 | 1 | 1 | 1 | 1 |
| 203 | 1 | 1 | 0 | 0 | 1 | 0 | 1 | 0 | 1 | 1 | 1 | 1 | 1 | 1 | 1 | 1 | 1 | 1 | 1 | 1 |
| 204 | 1 | 1 | 0 | 0 | 1 | 0 | 1 | 1 | 1 | 1 | 1 | 1 | 1 | 1 | 1 | 1 | 1 | 1 | 1 | 1 |
| 205 | 1 | 1 | 0 | 0 | 1 | 1 | 0 | 0 | 1 | 1 | 1 | 1 | 1 | 1 | 1 | 1 | 1 | 1 | 1 | 1 |
| 206 | 1 | 1 | 0 | 0 | 1 | 1 | 0 | 1 | 1 | 1 | 1 | 1 | 1 | 1 | 1 | 1 | 1 | 1 | 1 | 1 |
| 207 | 1 | 1 | 0 | 0 | 1 | 1 | 1 | 0 | 1 | 1 | 1 | 1 | 1 | 1 | 1 | 1 | 1 | 1 | 1 | 1 |
| 208 | 1 | 1 | 0 | 0 | 1 | 1 | 1 | 1 | 1 | 1 | 1 | 1 | 1 | 1 | 1 | 1 | 1 | 1 | 1 | 1 |
| 209 | 1 | 1 | 0 | 1 | 0 | 0 | 0 | 0 | 1 | 1 | 1 | 1 | 0 | 0 | 0 | 0 | 0 | 0 | 0 | 0 |
| 210 | 1 | 1 | 0 | 1 | 0 | 0 | 0 | 1 | 1 | 1 | 1 | 1 | 1 | 0 | 0 | 0 | 1 | 0 | 0 | 0 |
| 211 | 1 | 1 | 0 | 1 | 0 | 0 | 1 | 0 | 1 | 1 | 1 | 1 | 1 | 1 | 0 | 0 | 1 | 1 | 0 | 0 |
| 212 | 1 | 1 | 0 | 1 | 0 | 0 | 1 | 1 | 1 | 1 | 1 | 1 | 1 | 1 | 0 | 0 | 1 | 1 | 0 | 0 |
| 213 | 1 | 1 | 0 | 1 | 0 | 1 | 0 | 0 | 1 | 1 | 1 | 1 | 1 | 0 | 1 | 0 | 1 | 0 | 1 | 0 |
| 214 | 1 | 1 | 0 | 1 | 0 | 1 | 0 | 1 | 1 | 1 | 1 | 1 | 1 | 0 | 1 | 0 | 1 | 0 | 1 | 0 |
| 215 | 1 | 1 | 0 | 1 | 0 | 1 | 1 | 0 | 1 | 1 | 1 | 1 | 1 | 1 | 1 | 0 | 1 | 1 | 1 | 0 |
| 216 | 1 | 1 | 0 | 1 | 0 | 1 | 1 | 1 | 1 | 1 | 1 | 1 | 1 | 1 | 1 | 0 | 1 | 1 | 1 | 0 |
| 217 | 1 | 1 | 0 | 1 | 1 | 0 | 0 | 0 | 1 | 1 | 1 | 1 | 1 | 1 | 1 | 1 | 1 | 1 | 1 | 1 |
| 218 | 1 | 1 | 0 | 1 | 1 | 0 | 0 | 1 | 1 | 1 | 1 | 1 | 1 | 1 | 1 | 1 | 1 | 1 | 1 | 1 |
| 219 | 1 | 1 | 0 | 1 | 1 | 0 | 1 | 0 | 1 | 1 | 1 | 1 | 1 | 1 | 1 | 1 | 1 | 1 | 1 | 1 |
| 220 | 1 | 1 | 0 | 1 | 1 | 0 | 1 | 1 | 1 | 1 | 1 | 1 | 1 | 1 | 1 | 1 | 1 | 1 | 1 | 1 |
| 221 | 1 | 1 | 0 | 1 | 1 | 1 | 0 | 0 | 1 | 1 | 1 | 1 | 1 | 1 | 1 | 1 | 1 | 1 | 1 | 1 |
| 222 | 1 | 1 | 0 | 1 | 1 | 1 | 0 | 1 | 1 | 1 | 1 | 1 | 1 | 1 | 1 | 1 | 1 | 1 | 1 | 1 |
| 223 | 1 | 1 | 0 | 1 | 1 | 1 | 1 | 0 | 1 | 1 | 1 | 1 | 1 | 1 | 1 | 1 | 1 | 1 | 1 | 1 |
| 224 | 1 | 1 | 0 | 1 | 1 | 1 | 1 | 1 | 1 | 1 | 1 | 1 | 1 | 1 | 1 | 1 | 1 | 1 | 1 | 1 |
| 225 | 1 | 1 | 1 | 0 | 0 | 0 | 0 | 0 | 1 | 1 | 1 | 1 | 0 | 0 | 0 | 0 | 0 | 0 | 0 | 0 |
| 226 | 1 | 1 | 1 | 0 | 0 | 0 | 0 | 1 | 1 | 1 | 1 | 1 | 1 | 0 | 0 | 0 | 1 | 0 | 0 | 0 |
| 227 | 1 | 1 | 1 | 0 | 0 | 0 | 1 | 0 | 1 | 1 | 1 | 1 | 1 | 1 | 0 | 0 | 1 | 1 | 0 | 0 |
| 228 | 1 | 1 | 1 | 0 | 0 | 0 | 1 | 1 | 1 | 1 | 1 | 1 | 1 | 1 | 0 | 0 | 1 | 1 | 0 | 0 |
| 229 | 1 | 1 | 1 | 0 | 0 | 1 | 0 | 0 | 1 | 1 | 1 | 1 | 1 | 0 | 1 | 0 | 1 | 0 | 1 | 0 |
| 230 | 1 | 1 | 1 | 0 | 0 | 1 | 0 | 1 | 1 | 1 | 1 | 1 | 1 | 0 | 1 | 0 | 1 | 0 | 1 | 0 |
| 231 | 1 | 1 | 1 | 0 | 0 | 1 | 1 | 0 | 1 | 1 | 1 | 1 | 1 | 1 | 1 | 0 | 1 | 1 | 1 | 0 |
| 232 | 1 | 1 | 1 | 0 | 0 | 1 | 1 | 1 | 1 | 1 | 1 | 1 | 1 | 1 | 1 | 0 | 1 | 1 | 1 | 0 |
| 233 | 1 | 1 | 1 | 0 | 1 | 0 | 0 | 0 | 1 | 1 | 1 | 1 | 1 | 1 | 1 | 1 | 1 | 1 | 1 | 1 |
| 234 | 1 | 1 | 1 | 0 | 1 | 0 | 0 | 1 | 1 | 1 | 1 | 1 | 1 | 1 | 1 | 1 | 1 | 1 | 1 | 1 |
| 235 | 1 | 1 | 1 | 0 | 1 | 0 | 1 | 0 | 1 | 1 | 1 | 1 | 1 | 1 | 1 | 1 | 1 | 1 | 1 | 1 |
| 236 | 1 | 1 | 1 | 0 | 1 | 0 | 1 | 1 | 1 | 1 | 1 | 1 | 1 | 1 | 1 | 1 | 1 | 1 | 1 | 1 |
| 237 | 1 | 1 | 1 | 0 | 1 | 1 | 0 | 0 | 1 | 1 | 1 | 1 | 1 | 1 | 1 | 1 | 1 | 1 | 1 | 1 |
| 238 | 1 | 1 | 1 | 0 | 1 | 1 | 0 | 1 | 1 | 1 | 1 | 1 | 1 | 1 | 1 | 1 | 1 | 1 | 1 | 1 |
| 239 | 1 | 1 | 1 | 0 | 1 | 1 | 1 | 0 | 1 | 1 | 1 | 1 | 1 | 1 | 1 | 1 | 1 | 1 | 1 | 1 |
| 240 | 1 | 1 | 1 | 0 | 1 | 1 | 1 | 1 | 1 | 1 | 1 | 1 | 1 | 1 | 1 | 1 | 1 | 1 | 1 | 1 |
| 241 | 1 | 1 | 1 | 1 | 0 | 0 | 0 | 0 | 1 | 1 | 1 | 1 | 0 | 0 | 0 | 0 | 0 | 0 | 0 | 0 |
| 242 | 1 | 1 | 1 | 1 | 0 | 0 | 0 | 1 | 1 | 1 | 1 | 1 | 1 | 0 | 0 | 0 | 1 | 0 | 0 | 0 |
| 243 | 1 | 1 | 1 | 1 | 0 | 0 | 1 | 0 | 1 | 1 | 1 | 1 | 1 | 1 | 0 | 0 | 1 | 1 | 0 | 0 |
| 244 | 1 | 1 | 1 | 1 | 0 | 0 | 1 | 1 | 1 | 1 | 1 | 1 | 1 | 1 | 0 | 0 | 1 | 1 | 0 | 0 |
| 245 | 1 | 1 | 1 | 1 | 0 | 1 | 0 | 0 | 1 | 1 | 1 | 1 | 1 | 0 | 1 | 0 | 1 | 0 | 1 | 0 |
| 246 | 1 | 1 | 1 | 1 | 0 | 1 | 0 | 1 | 1 | 1 | 1 | 1 | 1 | 0 | 1 | 0 | 1 | 0 | 1 | 0 |
| 247 | 1 | 1 | 1 | 1 | 0 | 1 | 1 | 0 | 1 | 1 | 1 | 1 | 1 | 1 | 1 | 0 | 1 | 1 | 1 | 0 |
| 248 | 1 | 1 | 1 | 1 | 0 | 1 | 1 | 1 | 1 | 1 | 1 | 1 | 1 | 1 | 1 | 0 | 1 | 1 | 1 | 0 |
| 249 | 1 | 1 | 1 | 1 | 1 | 0 | 0 | 0 | 1 | 1 | 1 | 1 | 1 | 1 | 1 | 1 | 1 | 1 | 1 | 1 |
| 250 | 1 | 1 | 1 | 1 | 1 | 0 | 0 | 1 | 1 | 1 | 1 | 1 | 1 | 1 | 1 | 1 | 1 | 1 | 1 | 1 |
| 251 | 1 | 1 | 1 | 1 | 1 | 0 | 1 | 0 | 1 | 1 | 1 | 1 | 1 | 1 | 1 | 1 | 1 | 1 | 1 | 1 |
| 252 | 1 | 1 | 1 | 1 | 1 | 0 | 1 | 1 | 1 | 1 | 1 | 1 | 1 | 1 | 1 | 1 | 1 | 1 | 1 | 1 |
| 253 | 1 | 1 | 1 | 1 | 1 | 1 | 0 | 0 | 1 | 1 | 1 | 1 | 1 | 1 | 1 | 1 | 1 | 1 | 1 | 1 |
| 254 | 1 | 1 | 1 | 1 | 1 | 1 | 0 | 1 | 1 | 1 | 1 | 1 | 1 | 1 | 1 | 1 | 1 | 1 | 1 | 1 |
| 255 | 1 | 1 | 1 | 1 | 1 | 1 | 1 | 0 | 1 | 1 | 1 | 1 | 1 | 1 | 1 | 1 | 1 | 1 | 1 | 1 |
| 256 | 1 | 1 | 1 | 1 | 1 | 1 | 1 | 1 | 1 | 1 | 1 | 1 | 1 | 1 | 1 | 1 | 1 | 1 | 1 | 1 |